# Flow cytometry with anti-diffraction light sheet (ADLS) by spatial light modulation

*Yanyan Gong,\*,a Ming Zeng ,\*,a Yueqiang Zhu,\*,a Shangyu Li, a Wei Zhao, a, † Ce Zhang, a, †, Tianyun Zhao, b Kaige Wang, a Jiangcun Yang, c Jintao Bai a* a State Key Laboratory of Photon-Technology in Western China Energy, International Collaborative Center on Photoelectric Technology and Nano Functional Materials, Institute of Photonics & Photon Technology, Northwest University, Xi'an 710127, China.
b School of Automation, Northwestern Polytechnical University, Xi'an 710072, China.
c Department of Transfusion Medicine, Shaanxi Provincial People's Hospital, Xi'an 710068, China
\*These authors contribute equally to this investigation
† Correspondence: Wei Zhao: zwbayern@nwu.edu.cn; Ce Zhang: zhangce@nwu.edu.cn

**Abstract**

Flow cytometry is a widespread and powerful technique, whose resolution is determined by its capacity to accurately distinguish fluorescently positive populations from negative ones. However, most informative results are discarded while performing the measurements of conventional flow cytometry, e.g., the cell size, shape, morphology, and distribution or location of labeled exosomes within the unpurified biological samples. We, herein, propose a novel approach using an anti-diffraction light sheet with anisotroic feature to excite fluorescent tags. Constituted by an anti-diffraction Bessel-Gaussian beam array, the light sheet is 12 µm wide, 12 µm high, with a thickness of ~ 0.8 µm. The intensity profile of the excited fluorescent signal can, therefore, reflect the size and allow samples in the range from O(100 nm) to 10 µm (e.g., blood cells) to be transported via hydrodynamic focusing in a microfluidic chip. The sampling rate is 500 kHz provides a capability of high throughput without sacrificing the spatial resolution. Consequently, the proposed anti-diffraction light-sheet flow cytometry (ADLSFC) can obtain more informative results than the conventional methodologies, and is able to provide multiple characteristics (e.g., the size and distribution of fluorescent signal) helping to distinguish the target samples from the complex backgrounds.

## 1. Introduction

Flow cytometry (FC) is a powerful analytical technique that enables rapid analysis of cells and particles in solutions flowing past a single or multiple laser intercept points(McKinnon, 2018). Due to its ability to count, characterize and sort cells, it has been widely used in cell analysis and disease diagnosis(Barteneva et al., 2012; Elshal & McCoy, 2006; Levine et al., 2021; Mei et al., 2016).

In the past nearly three decades, several research groups around the world have been involved in the study of microfluidic flow cytometry(de Rutte et al., 2022; Dittrich et al., 2006; Erdem et al., 2020; He et al., 2014; Liascukiene et al., 2018; Martel & Toner, 2013; Wang et al., 2013). There are various types of flow cytometers have been developed, including acoustic focusing cytometers, cell sorters, imaging cytometers, mass cytometers, and cytometers for bead array analysis(McKinnon, 2018). Acoustic focusing cytometry uses ultrasound to help focus cells for laser interrogation(Ward et al., 2009). The advantage is that it does not require a high speed or large volume of sheath flow. However, it suffers from low throughput and complex and large equipment. The cell sorter separates cells by generating droplets with high-frequency oscillation of the liquid sample stream(Webster et al., 1988). The droplets are then given a positive or negative charge and passed through a metal deflector plate. Eventually, they are directed to specific collection containers depending on their charge. The cell sorter has high separation purity and flexibility, but it has no sample detail information.

In combination with fluorescence microscopy, imaging flow cytometry (IFC) (Heidi et al., 2007) allows rapid analysis of morphology and multi-parameter fluorescence of biological samples at the single cell level. While, a high spatial resolution of IFC can only be achieved by sacrificing the throughput, i.e., the number of analyzed biological samples per second. For example, the time span required for imaging one cell using a charged coupled device (CCD) can be as short as 1 ms, which means a maximum throughput of only 1000 cells per second(Mikami et al., 2020). Also, IFC requires large storage space and high cost of analyzing time.

Mass cytometry combines time-of-flight mass spectrometry and flow cytometry(Matthew et al., 2016). Cells are labeled with heavy metal ion-labeled antibodies (usually from the lanthanide family) instead of fluorescent antibodies and detected using time-of-flight mass spectrometry. Since fluorescent labeling is not used, no light compensation is required. But the sample cells are destroyed and cannot be analyzed downstream. The cytometer for bead array analysis is a technique that detects the fluorescent beads specifically bonding to the samples. By evaluating the intensity of fluorescence, the number of samples associated with the beads can be quantified(Scorzetti et al., 2009).

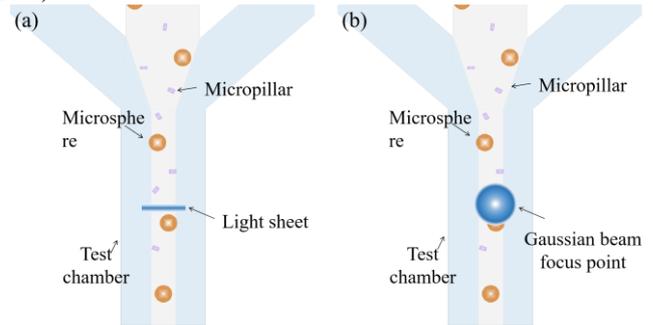

Figure 1. (a) Fluorescent micropillars and microspheres pass through light sheet, and (b) Fluorescent micropillars and microspheres pass through a Gaussian beam spot.

Conventional flow cytometry is usually "bulky" and has relatively low separation purity(McKinnon, 2018). With the fast development of microfluidics and miniature techniques, portable, highly integrated, and easy-to-operate flow systems have been developed in the last ten years(Han et al., 2014; Leipold et al., 2015; Ward & Kaduchak, 2018). For example, Jiang et al. designed a miniaturized dual-wavelength fluorescence detection chip by coupling electrokinetically induced pressure-driven flow for fluorescent particle counting(Jiang et al., 2013). Kanwa et al. used polydimethylsiloxane (PDMS) to fabricate a microfluidic device shaped like a gourd to capture and quantitatively analyze stained exosomes(Kanwar et al., 2014). Lee et al. proposed an acoustic nano filter system that separates microvesicles at specific sizes in a continuous and contactless manner(Lee et al., 2015).
These new cell (or particle) detection devices have enriched the family of flow cytometry.

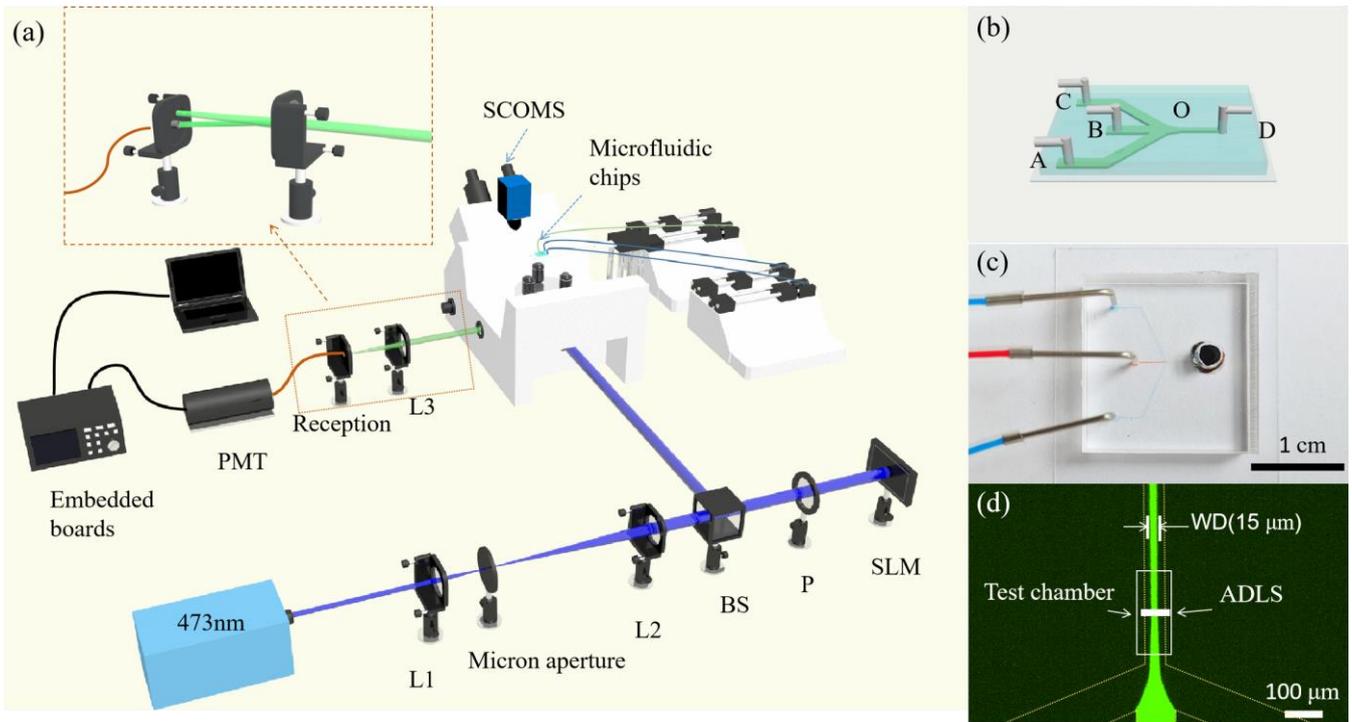

Figure 2. Schematic diagram of the ADLSFC system. (a) Schematic of the system. (b) microfluidic chip, where A and C are the inflow port of sheath flow, B is the sample inlet, and D is the outlet. (c) Photo of the microfluidic chip. (d) fluorescence image of the microchannel. The inlets A and C were injected with pure water, and B was injected with fluorescein sodium salt solution. All inlets have flow rates of 1 μL/h. The width of the jet (WD) is 15 μm.

Nevertheless, the performance of FC is the compromise of one of the key parameters in favor of the others, hence limiting its applications. Morphological features of the biological samples are often among the key parameters, which help to distinguish one population from another(Mikami et al., 2020). For example, escherichia coli is a typical gram-negative rod bacterium, whose length ranges from 1 to tens of micrometers and reflects distinctive activities(Tans et al.). In the conventional FCs where elliptical and spherical focal spots are adopted, the differences in the size, shape and orientation of cells when passing through intercept points can not be detected, which makes it difficult to distinguish the active cells (Zhao et al., 2016). In contrast, focal spots with a large aspect ratio (e.g., the light sheet) generate distinctive intensity profile when a non-spherical object passes (Fig. 1). Taichi Miura etc. (Taichi et al., 2018) implement a light sheet into a customized IFC, which aims to boost the fluorescence intensity at each image pixel by a factor of ~10. To the best of our knowledge, none of these researches utilize focal spot with high aspect ratio to improve spatial resolution of FC.

In this investigation, we designed a novel flow cytometry system using a highly anisotropic and anti-diffraction light sheet (ADLS) as the laser interception point. The ADLS was formed by a parallelly and tightly aligned Bessel-Gaussian beam generated by a spatial light modulator (SLM) with stripe split phase (SSP) method (Zhu et al., 2021). Dimension of the ADLS is 12 μm by 12 μm, and 0.8 μm in thickness. Passing through objects with different sizes results in distinctive intensity profiles which can be counted and analyzed. Coupled with a photomultiplier tube and high-speed data acquisition module, a 500 kHz sampling rate can be easily achieved. The theoretical throughput can be at least $10^4$ events per second. Therefore, the proposed methodology greatly improves the spatial resolution of conventional FC, and is suitable for sorting and analysing biological samples with complex physical and chemical features.

## 2. System setup

The experimental system is shown in Figure 2(a). In this experiment, a 473 nm continuous wave (cw) laser (MW-RL-473, 200 mW, CNI) is adopted as the light source. The laser beam is first filtered by a spatial light filter and then collimated by a beam expander. The beam is further adjusted through a beam splitter (THORLABS, BS 400 nm ~ 800 nm) and a polarizer (P). After passes through the polarizer, the direction of the linearly polarized beam is parallel to that of the liquid crystal (LC) of the SLM (LETO, HOLOEYE Photonics AG, Germany, PLUTO-NIR-011, 420 nm ~ 1100 nm). The modulated beam, i.e., ADLS, is further directed into an inverted fluorescence microscope system (NIB900, NEXCOPE, China) and focused with an objective (Leica, PL Apo 20X, NA 0.4) to generate an ADLS at the focus. The microscope has a set of filters, including a dichroic mirror (SEMROCK, Di01-R488/543/635, 473 nm HR) and a band-pass filter (CHROMA, ZET488/640 nm) to extract fluorescence from the excitation light.

Particle detection, measurement, and analysis are achieved using the ADLS. Accompanied, a microfluidic chip was fabricated for flow cytometry (Figure 2(b, c)). The microchip has three inlets. The A and C inlets are for water flow, while the B inlet is for the sample solution. The three solutions contact at the joint O and form a sheath flow downstream. The microchannels AO, BO, and CO are 100 μm in width and 10 μm in height. The test chamber, i.e., the DO section, is 50 μm in width and 10 μm in height. The ADLS is placed in the test chamber to detect the sample in the sheath flow, as shown in Figure 2(d).

The spatial position of the microfluidic chip is changed by a nano-piezo stage (CB7 4EX, THORLABS, UK) so that locating the ADLS in the center of the microchannel. When the fluorescent particles pass through the ADLS at a given speed, the ADLS excites the fluorescent particles to emit fluorescence. The fluorescent signal passes through the band-pass filter and dichroic mirror in turn, then is captured by detectors. With a beam splitter in the microscope, 20% of the fluorescence is captured by a SCMOS camera (PCO edge 4.2LT, Kelheim, Germany, PCO) to monitor the ADLS. The remaining 80% fluorescence is focused into an optical fiber and detected by a photomultiplier tube (Hamamatsu Photonics, H7415, 300~650 nm) (PMT). As shown in the inset of Figure 7(a), only the fluorescence from the modulated ADLS is focused into the optical fiber. The possible fluorescence excited by the zeroth order light has been blocked.

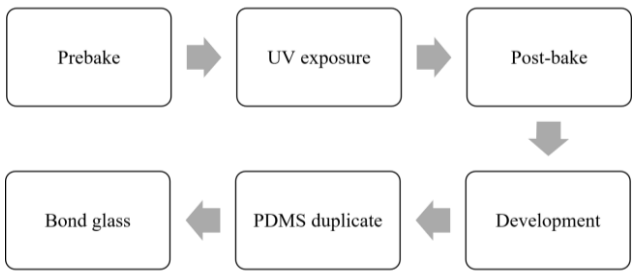

Figure 3. Fabrication process of the microfluidic chip.

The PMT, combined with an amplifier (low gain x$10^5$, medium gain x$10^6$, and high gain x$10^7$), converts fluorescence signals into electrical signals. Then, after collected and converted into digital raw data by a 16 bits analogue-to-digital converter on an embedded board, the data are finally sent to a computer for processing and analysis. The sampling rate of the system is up to 500 kHz.

*Fabrication of the microfluidic chip*

The microchannel is fabricated through soft-lithography with polydimethylsiloxane (PDMS). The channel layer was first fabricated by a negative photoresist SU-8 3025 on a silicon wafer. After 10 min prebake, the photoresist-coated wafer is exposed to UV light using a contact mask aligner (MIDAS, MDA400LJ, Korea). After post-baking (10 min) and development, the template is obtained. The structure of the template is subsequently duplicated by PDMS. The PDMS microchannel is further bonded onto a glass slide and punched with inlets and outlet to form the desired microfluidic chip.

*Sample preparation*

To test the ADLSFC system, we used two kinds of fluorescent particles: polystyrene fluorescent micropillar (0.4~1 μm long with 0.3~0.6 μm diameter) and 5 μm diameter polystyrene fluorescent microspheres. Both of them have an excitation peak at 468 nm and an emission peak at 508 nm. Thermo Fisher Scientific, USA) (Figure 4(a, b)). Both microparticles were diluted with ultrapure water. To prevent the fluorescent beads from sticking to the wall of the microchannel, surfactant of Pluronic F-127 (Merck, German) was added to the diluted solution, and the final concentration of Pluronic F-127 was less than 0.1%.

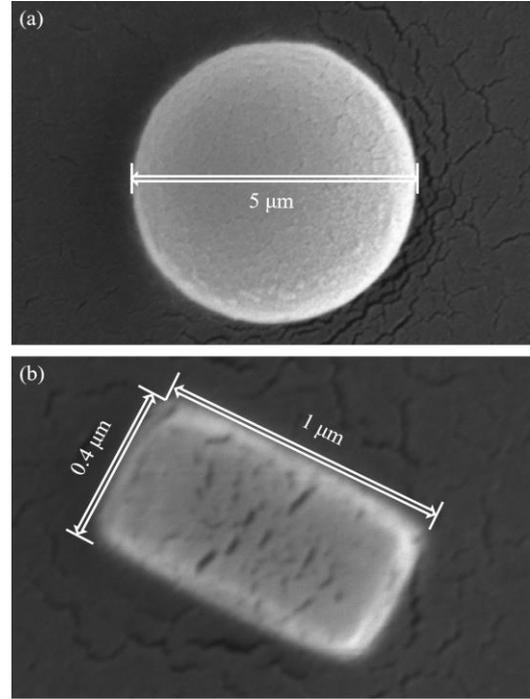

Figure 4. Scanning electron microscope (SEM) image of fluorescent particles. (a) Fluorescent microspheres. (b) A typical fluorescent micropillar.

In the experiment, we used two syringe pumps to provide the basic flow and deliver the microparticles. Three 50 μL syringes were connected to the inlet A, B, and C of the microfluidic chip, with peek tubing and connectors. Water was injected into inlet A and C to provide sheath flow. Fluorescent microparticles were injected into inlet B. Before measurement, the microchannel was degassed in advance. The detection was carried out right after the sheath flow was formed stably. The particle velocities in the jet at different flow rates ($Q$) were obtained using a camera, as plotted in Figure 5. The jet-to-sheath flow rate ratio is 1:1. For each flow rate, the particle velocity measurement is repeated 20 times. From Figure 5, it can be seen the higher the given flow rate, the faster the particle velocity, accompanied with a higher velocity fluctuation. When $Q$ is less than 1 μL/h, a stable sheath flow cannot be formed in the test chamber. Therefore, we choose 1 μL/h as the experimental flow rate, and the average velocity ($v$) of particles is 1.41 mm/s. The jet width (WD) in the test chamber is 15 μm.

*Data analysis*

With a set flow rate, the speed of particle can be calculated
$$v = kQ/A \qquad (1)$$
$k$ is a correction coefficient, $A$ is the microchannel cross-sectional area. Then, the diameter ($d_p$) of the particle can be estimated as
$$d_p = v\tau - d_s \qquad (2)$$
$d_s$ is the thickness of the ADLS, $\tau$ is the full width at half maximum (FWHM) of the pulse signal in time domain corresponding to the particle.

In summary, after determining $\tau$ of the particle, the particle size can be obtained directly relying on the particle speed.

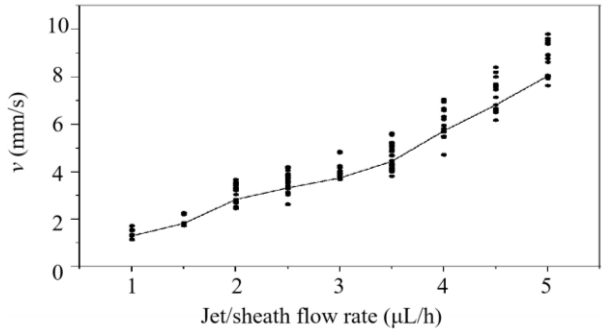

Figure 5. Fluorescent particle velocity varies with jet/sheath flow rate. The jet-to-sheath flow rate ratio is 1:1. Each dot represents a measurement. The average velocity of the measurement is represented by the solid line.

The data processing algorithm mainly includes five steps, as diagrammed in Figure 6. (1) The raw data is first processed by a threshold filter and a bandpass filter to remove noise and pick the pulse regions from the raw data. (2) Smooth these pulse regions with a moving average algorithm and search for the local maxima in the pulse regions. (3) Determine the widths of the pulses from their rising and falling edges, then, validate the pulse data with the bandpass filter again. (4) Interpolate the validated pulse data to calculate the $\tau$. (5) Repeat the process above and statistically calculate the number and diameters of the particles.

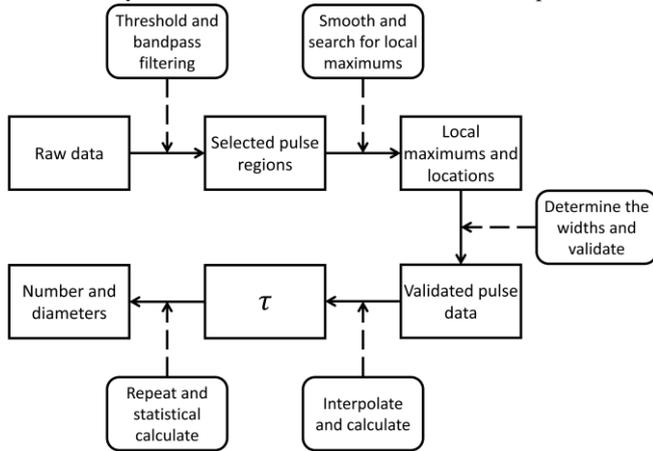

Figure 6. Diagram of data analysis.

## 3. Generation of ADLS

Conventional flow cytometers use a focused spot as the laser interrogation point. The axisymmetrically distributed focus spot cannot completely cover the microchannel interface, in the meanwhile obtaining the detailed information about the sample. In contrast, a highly anisotropic ADLS can cover the entire microchannel with a large area, in the meanwhile, reserve a small thickness to provide high resolution in the streamwise direction.

There are two reported ways of generating a light sheet, including (1) using a cylindrical lens and (2) by rapidly scanning the beam back and forth (Keller et al., 2010; Keller et al., 2008; Planchon et al., 2011). The formation of a light sheet through a cylindrical lens is easy to integrate into the optical path because it does not require moving parts, but the thickness of the light sheet is normally 2~10 μm(Girkin & Carvalho, 2018; Hedde et al., 2016; Heddleston & Chew, 2016; Voie et al., 2011).

The fast-scanning beam method does not really generate a light sheet, and it is more suitable for detecting stationary samples. It could lose the samples in the detection of the flow system.

Relative to the methods above, a SLM that can be easily integrated into the optical system to generate a thin light sheet as the laser intercept point is very suitable for developing novel flow cytometry. By SLM, diverse laser spots with desired properties, e.g., anti-diffraction beam, self-bending beam, and negative refractive index beams etc., can be generated as demand. Since the anti-diffraction feature of Bessel-Gaussian (BG) beam(Richards et al., 1959), numerous BG beams can be aligned sufficiently close to each other to form a light sheet (Fig. 7a), i.e., ADLS, without significant influence of interference. BG beam has a smaller diameter of focus than a Gaussian beam, but a much larger depth of focus. Therefore, the ADLS constituted of BG beams simultaneously reserve large areas with small thicknesses.

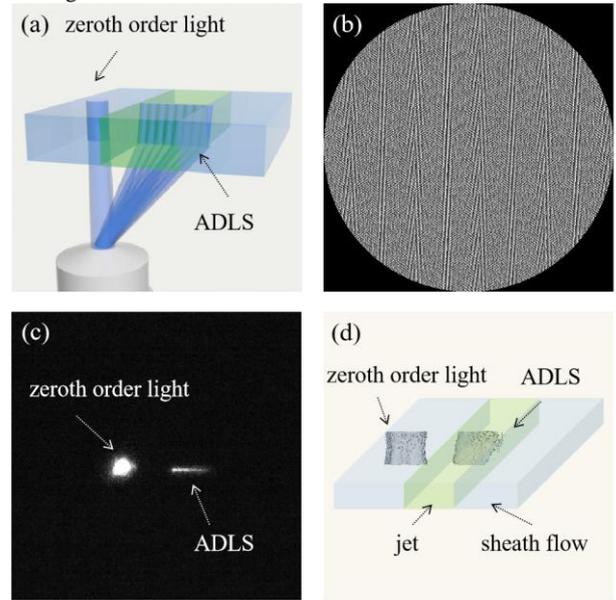

Figure 7. (a) Schematic diagram of the generation of ADLS at the focal point of the lens. Zeroth order light spots are separated with the modulated ADLS. The latter is located in the center of the jet, while the former is located in the sheath flow. (b) The phase map of generating the ADLS. (c) Experimental visualization of the zeroth order light and ADLS by an inverted fluorescence microscope, using a 50 μmol/L concentration fluorescein sodium salt (SIGMA, 46970-100G, Germany) solution. (d) 3D construction of the ADLS from fluorescent images by MATLAB software. The ADLS is 12 μm wide, 12 μm high, and 0.8 μm thick.

The BG beam and the ADLS generated by SLM can be elucidated by Debye diffraction theory(Leutenegger et al., 2006; Richards et al., 1959). Based on Fourier transform (FT) of the Debye diffraction integral and ignoring constant coefficients, the electric field intensity ($E$) of light can be calculated by (Zhu et al., 2021)

$$E(x,y,z) = \begin{bmatrix} E_x \\ E_y \\ E_z \end{bmatrix} = \int_{-q}^{q}\int_0^{2\pi} \frac{p(\theta)E_t(\theta,\varphi)}{\cos\theta} e^{ik_z z} e^{i(k_x x + k_y y)} dk_x dk_y$$
$$= F^{-1}[p(\theta)E_t(\theta,\varphi)e^{ik_z z}/\cos\theta ] \quad (3)$$

where $p(\theta)$ is the apodization function of the objective lens, $E_t(\theta,\varphi)$ is the transmission electric field, $\varphi$ is the azimuth angle of the objective lens, $\theta = \arcsin(rNA/Rn_t)$ is the convergence angle of the objective lens, and $k_x, k_y, k_z$ are the wavenumbers in vacuum. Here, $F^{-1}$ denotes inverse Fourier transform.

According to Eq. (3), the electric field of modulated beam is

$$E(x,y,z) = F^{-1}\{[e^{i\Phi}]P(\theta)E_t(\theta,\varphi)e^{ik_z z}/\cos\theta\} \quad (4)$$

where $\Phi$ is the modulation phase function. In this investigation, to generate Bessel-Gaussian beam with blazed grating simultaneously, we have

$$\Phi = \Phi_1 + \Phi_2 \quad (5)$$

$\Phi_1$ and $\Phi_2$ are the phase functions for modulating blazed grating and Bessel-Gaussian beams. They can be expressed as(Zhu et al., 2021)

$$\Phi_1(x', y') = \frac{2\pi}{\lambda}\left(x'\,\Delta x + y'\,\Delta y\right) \quad (6)$$
$$\Phi_2 = 2\pi r \tan \alpha / \lambda \quad (7)$$

where $x', y'$ be the Cartesian coordinates in the pupil plane, NA is the numerical aperture of the objective lens, $R$ is the maximum radius of the pupil plane of the objective lens, $n_t$ is the refractive index of the objective. $r = \sqrt{x'^2 + y'^2}$ is the polar diameter in the plane of the pupil plane, $\alpha$ is the apex angle of the angular pyramid prism, $\lambda$ is the wavelength. Finally, the light intensity after modulation can be calculated through $I = |E|^2$.

Given $\Delta x, \Delta y$ for each Bessel-Gaussian beam, the ADLS formed by aligned Bessel-Gaussian beams can be generated, as shown in Figure 7(a). To achieve a uniform ADLS in the experiment, nine Bessel-Gaussian beams were generated and arranged in parallel at an interval of 8λ. The BG array is generated via a SLM, with the phase map designed according to strip segmentation phase (SSP) method(Zhu et al., 2021), with a blazed grating superposed (Figure 7b). In order to evaluate the ADLS directly, a thin layer of fluorescein sodium salt solution (SIGMA, 46970-100G, Germany) with 50 μmol/L concentration was spin-coated solution a coverslip as calibration film. Its thickness is 2 μm. By moving the calibration film vertically, the structure of ADLS is visualized by a SCMOS camera layer-by-layer. From Figure 7(c), it can be seen the zeroth order light has been distantly moved from the modulated beams, showing negligible influence on the fluorescent measurement. The 3D distribution of ADLS was reconstructed by Matlab software, as shown in Figure 7(d). The ADLS is 12 μm wide, 12 μm high and $d_s$ =0.8 μm thick if a 20X NA 0.4 lens was applied. It is sufficiently large to cover the entire sheath flow.

## 4. Results

The performance of the ADLSFC system is evaluated by the fluorescent micropillars and microspheres, respectively. Then, the mixture of the micropillars and microspheres is applied in the experiments to show the capability of distinguishing different samples in a wide size range.

*Detection of the micropillars*

First, the fluorescent micropillar with a relatively high concentration ($\sim 10^7$ mL$^{-1}$) is applied. A typical time trace of the fluorescence signal ($I_f$) is plotted in Figure 8(a) for the measurement of fluorescent micropillars. It can be seen the raw fluorescent signal contains a large amount of noise, including optical noise, instrumental thermal noise, and electromagnetic noise. The presence of noisy signals can generate analysis errors on the FWHM of the pulse. Thus, a series of denoise methods have been applied to remove noise.

The result after processing is shown in Figure 8(b). Here, a medium gain was applied to amplify the weak fluorescent signal. It can be seen that the peak fluorescence ($I_{fp}$) of the micropillar signal is $1.38 \times 10^4$ and $\tau$ =1.13 ms. The diameter of the particle is 1.09 μm, as estimated from Eq. (2). With ADLSFC, micropillars with submicron sizes ($d_p =$ 0.73 μm and $d_p = 0.39$ μm) are also distinguished, as shown in Figure 8(c-f).

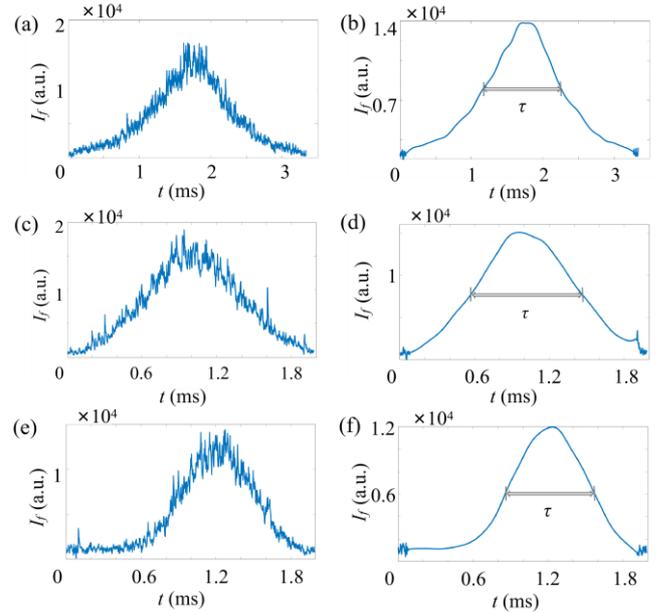

Figure 8. Typical time traces of fluorescence signals of micropillars in a low particle concentration, $\sim 10^7$ mL$^{-1}$. The horizontal axis represents the time, and the vertical axis represents the signal strength. Here, a moderate gain is applied for the PMT. (a, c, e) raw data and (b, d, f) data after noise reduction. (a, b) Time trace of micropillar with $d_p = 1.09$ μm. (c, d) Time trace of micropillar with $d_p = 0.73$ μm. (e, f) Time trace of micropillar with $d_p = 0.39$ μm.

Theoretically, the number of micropillars injected into the test chamber in 5 minutes is 833. Our measurement with the ADLSFC is 736 on average. The detection efficiency is $\delta = 88.3\%$. Since the theoretical particle number is estimated from the manufacturing value, the actual concentration of particles could be reduced according to many factors, e. g., sedimentation, and adhesion. Thus, the injected number of particles must be smaller than the theoretical value. Accordingly, the actual $\delta$ could be higher than 88.3%. The high detection efficiency reflects the high sensitivity of the system.

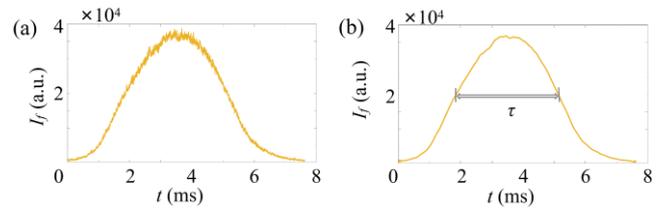

Figure 9. Typical time traces of fluorescence signals of microspheres in a low particle concentration, $\sim 8 \times 10^4$ mL$^{-1}$. The horizontal axis represents the time, and the vertical axis represents the signal strength. Here, a low gain is applied for the PMT. (a) raw data, (b) after noise reduction.

*Detection of the microspheres*

Then, the fluorescent microspheres is detected at a low concentration ($\sim 8 \times 10^4$ mL$^{-1}$), which commonly exists in biological and biomedical applications where large but sparse cells or bacteria targets are detected. A typical time trace of $I_f$ is plotted in Figure 9(a). The corresponding result after processing is shown in Figure 9(b). Since the fluorescence of the 5 μm microsphere is relatively strong, a low gain was applied in the measurement. It can be seen that $I_{fp}$ of the microsphere signal is

$3.68 \times 10^4$ and $\tau$ =3.46 ms. The diameter of the particle is 4.98 μm from Eq. (2). Theoretically, the number of microspheres injected into the test chamber in 5 minutes is 7. Our measurement with the ADLSFC is 7.3 on average. The detection efficiency is $\delta = 104.3\%$. The number of microspheres detected in the experiment is consistent with the theoretical one. The sensitivity of ADLSFC in detecting sparse targets is as expected.

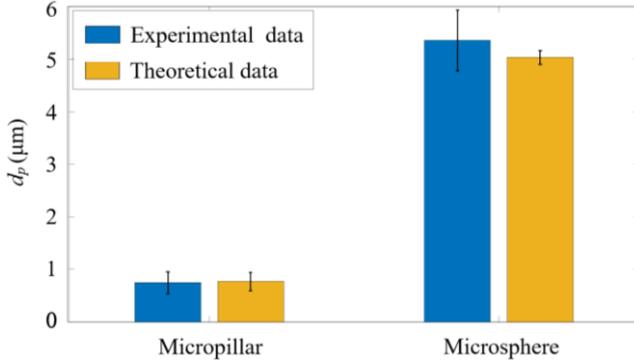

Figure 10. Experimental measurements versus theoretical particle size analysis. The blue histogram represents the size of the particles pass through ADLS. The theoretical data are estimated from SEM images of the particles.

### Size analysis of the fluorescent microspheres and micropillars

Subsequently, to verify the accuracy of ADLSFC for sample size measurement, we analyzed the measured particle size. As shown in Figure 10, the blue bars indicate the average diameters of micropillars and microspheres obtained by analyzing the FWHM of the fluorescence signals, and the yellow bars indicate the theoretical particle diameters. It can be seen the average $d_p$ of the micropillar from the experiments is 0.74 μm, while that of the theoretical one is 0.77 μm (estimated from the diagonal size of the micropillars measured in SEM). The results measured by ADLSFC are highly consistent with that by SEM. Thus, the validity of ADLSFC to detect and measure the size of submicron scale particles has been supported. In the meanwhile, the $d_p$ of the microsphere from the experiments is basically consistent with the theoretical one. All these indicate that the ADLSFC can reliably evaluate the streamwise size of the particles at the moment of passing through the ADLS.

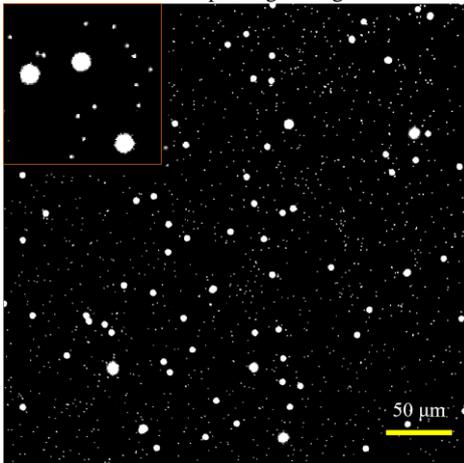

Figure 11. Fluorescence image of mixed micropillars and microspheres solution. The shape of the micropillar is not distinguishable from the image due to the low resolution.

### Detection of mixed microparticles with different proportions

Since the two fluorescent particles have significantly different size ranges that spanned a decade, the large fluorescent intensity difference may cause big trouble in detecting the micropillars and microspheres simultaneously. In the detection of micropillars and microspheres separately above, two different gains have been applied. However, in the detection of mixed particle solution, we have to use a single gain. Therefore, in this section, we show whether ADLSFC can distinguish the two particles from the mixed solution.

Before the experiment, the concentration of the particles and their ratio in the mixed solution is preliminarily determined by fluorescent image analysis, as shown in Figure 11. The micropillar solution is diluted 500 times, while the microsphere solution is diluted 60 times, and then mixed in equal volume. In theory, the particle density ratio ($\beta = C_{pi}/C_s$, where $C_{pi}$ and $C_s$ are the particle density of the micropillar and microsphere, respectively) between micropillars and microspheres is 15. The actual $\beta$ obtained from the fluorescent image is approximately 16.67. The measurement result by the image method is approximately consistent with the theoretical one.

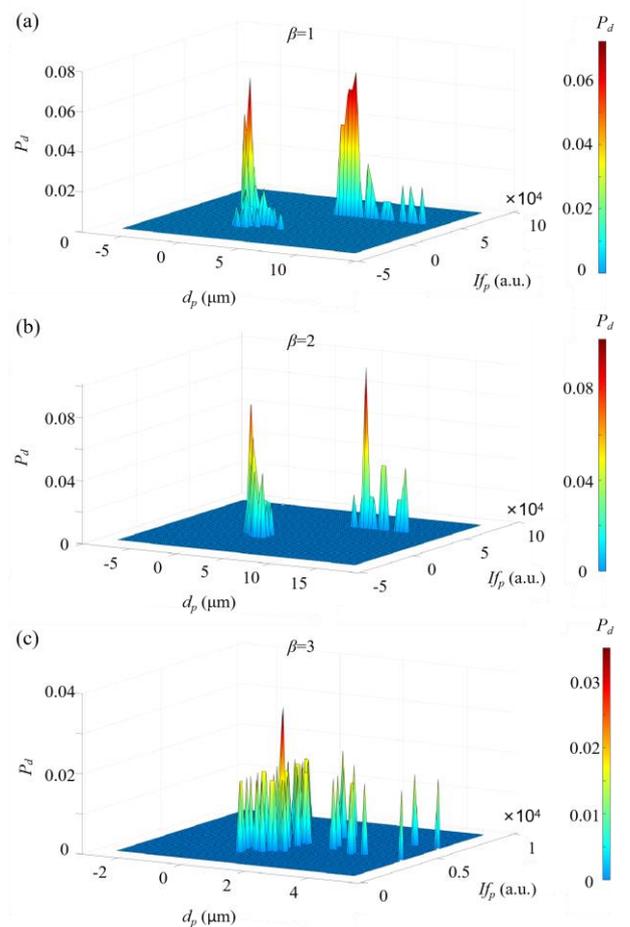

Figure 12. Two-dimensional probability density distribution of a mixed solution. (a) $\beta = 1$, (b) $\beta = 2$, and (c) $\beta = 3$.

In this investigation, the micropillars and microsphere stock solutions were diluted 12500 and 100 times, respectively. Theoretically, the diluted solution concentration was $\sim 8 \times 10^5$ mL$^{-1}$. The two diluted solutions were mixed in the ratio of $\beta = 1, 2$, and 3, respectively.

First of all, we explored the capability of ADLSFC to measure particle size and distinguish different particles in the mixed solution. During the experiments, a medium gain was applied. The results are demonstrated by a two-dimensional probability density ($P_d$) distribution in a two-parameter space, including the particle size $d_p$ and the peak fluorescence intensity $I_{fp}$, as shown in Figure 12. Normally, the larger the particle size, the stronger the fluorescence. For instance, in Figure 12(a) where $\beta$ is 1, it can be clearly seen that $P_d$ shows two different clusters in the parameter space. One is around $d_p$ =0.70 μm and the other is around $d_p$ =6.68 μm. The former is coincident with Figure 10, while the latter is slightly larger with occasional measurement extended to ~10 μm. This is because the medium gain leads to a saturation of fluorescent signal during analogue-to-digital conversion. Thus, $I_{fp}$ is underestimated with the FWHM of the microspheres overestimated. Even though, Figure 12(a) indicate the micropillars and microspheres can be clearly distinguished by their sizes. Similar results are also observed in Figure 12(b). In Figure 12(c), due to the larger concentration of micropillars, the agglomeration of particles leads to separate spikes in the probability density distribution. Two clusters of particles with different sizes and peak intensities can still be distinguishable. These results support the capability and reliability of measuring particle sizes and distinguishing the particles from their size information by ADLSFC.

The experimental observation on the ratios between micropillars and microspheres is compared with the theoretical ones, as shown in Figure 13. When the theoretical $\beta$ is 1, 2, and 3, the experimental ones are 1.039, 2.252, and 3.394, respectively.

Although the experimental results on $\beta$ are slightly higher than the theoretical ones, the average differences are within 10%. The small difference can be attributed to two reasons. One is the particle density in the solution may fluctuate with time, resulting in a fluctuation of the theoretical particle density. The other is the microspheres with larger size tend to precipitate at such a low flow rate. This leads to a higher ratio between the micropillars and microspheres. In general, the measured particle ratios are consistent with the actual ratios, and the system again exhibits sufficient accuracy in identifying and sorting fluorescent particles of different particle size

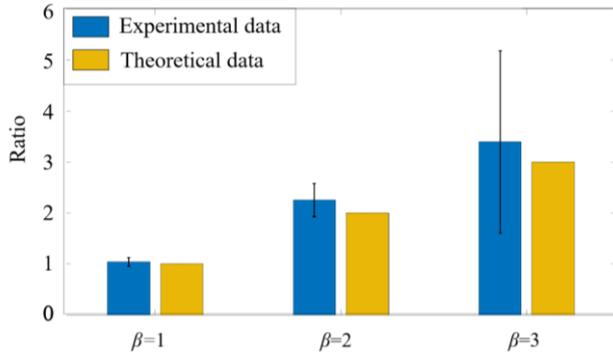

Figure 13. Comparison between the experimental ratios and the theoretical ones in the mixed solutions. The blue histogram represents the experimental measurement results, and the yellow histogram is used as a theoretical value comparison.

The 2D $P_d$ distribution provides not only the information on size distribution, but also intensity distribution information, which could imply the side area of the particle. Accordingly, two-dimensional information about an anisotropic sample, which might be important in biomedical and material investigations, can be revealed simultaneously.

*Analysis of the size distribution of fluorescent micropillars*

The micropillars have diverse sizes, with the length of 0.4~1 μm and the width of 0.3~0.6 μm. In the meanwhile, their attitudes when across the ADLS are unknown. Therefore, we analyzed the sizes of the micropillars when they pass through the ADLS to test the performance of ADLSFC on distinguishing submicron size particles. The results are shown in Figure 14.

Among the measured micropillars, their sizes are highly non-uniform. 30.6% of them are between 0.2-0.4 μm and show the highest portion (Figure 14, group a). In contrast, the micropillars between 0.8-1.0 μm have the lowest portion (8.3%) in the solution (Figure 14, group b). There are almost 25.4% of micropillars have a size over 1 μm and over half of the micropillars are below 0.4 μm. The successful detection of the small micropillars, as shown in both Figures 10 and 14, indicates the ADLSFC system is capable of capturing O(100 nm) particles, even using a 20X lens. The sensitivity and resolution of the system can be further improved with a high magnification and numerical aperture lens. Besides, if samples with regular sizes are applied, the system can even reveal the attitudes of the samples through their sizes when they pass through the ADLS.

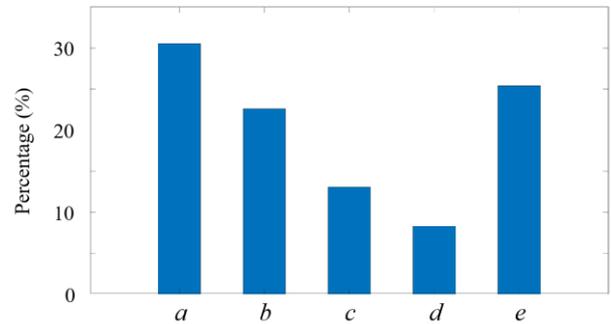

Figure 14. Percentage of fluorescent micropillars passing through ADLS. (a) 0.2 μm ≤ $d_p$ ≤ 0.4 μm, (b) 0.4 μm < $d_p$ ≤ 0.6 μm, (c) 0.6 μm < $d_p$ ≤ 0.8 μm, (d) 0.8 μm < $d_p$ ≤ 1 μm, (e) $d_p$ > 1 μm.

## Conclusions

In this research, we have developed a novel flow cytometry with an anti-diffraction light sheet. The light sheet is constituted of Bessel-Gaussian beams, which are parallelly and tightly aligned. With a large aspect ratio and small thickness, the ADLS can count and measure the sizes of particles and biological samples (e.g., cells) in a wide size range, from O(100 nm) to 10 μm. Thus, screening the biological samples by their sizes can be easily achieved.

By using commercial fluorescent particles, including both micropillars and microspheres, the performance of the ADLS flow cytometry is tested. The averaged particle detection efficiency is up to 96.3%. During the size analysis of particles in mixed solutions, the analytical error is within 10%. The analytical detection results match the theoretical values even at low particle density. All these fully demonstrate the good detection performance of ADLS flow cytometry, which could be an effective approach for micro/submicron-scale biological and particle material analysis and detection.

## Acknowledgements

This investigation is supported by National Natural Science Foundation of China (Grant No. 51927804, 61775181, 61378083).